\documentclass[prd,a4paper,nofootinbib,
showpacs, 
]{revtex4}
\usepackage{graphicx}
\usepackage{amsfonts}
\usepackage{slashed} 
\usepackage{amstext}
\usepackage{amsmath}
\def\eq#1{{eq.~(\ref{#1})}}

\def\hc{\mbox{h.c.}\,}

\def\gtap{\ \raisebox{-.6ex}{\rlap{$\sim$}} \raisebox{.4ex}{$>$}\ }

\def\hbar{\hspace{0pt}\raisebox{1pt}{$-$} \hspace{-7pt} h}

\def\5{\overline 5}

\newcommand{\be}{\begin{equation}}
\newcommand{\ee}{\end{equation}}
\newcommand{\bea}{\begin{eqnarray}}
\newcommand{\eea}{\end{eqnarray}}
\newcommand{\nn}{\nonumber}

\begin{document}
\title[]{Limits on  anomalous top quark gauge couplings from  Tevatron and  LHC data}
\date{\today}
\author{M. Fabbrichesi$^{\dag}$}
\author{M. Pinamonti$^{\ddag}$}
\author{A. Tonero$^{\circ}$}
\affiliation{$^{\dag}$INFN, Sezione di Trieste}
\affiliation{$^{\ddag}$INFN, Sezione di Trieste, Gruppo collegato di Udine and SISSA, via Bonomea 265, 34136 Trieste, Italy}
\affiliation{$^{\circ}$ ICTP-SAIFR \& IFT, Rua Dr.\ Bento Teobaldo Ferraz 271, 
01140-070 S\~ao Paulo, Brazil }
\begin{abstract}
\noindent  We review and update current limits on possible anomalous couplings of the top quark to gauge bosons. We consider data from top quark decay (as encoded in the $W$-boson helicity fractions) and single-top production (in the $t$-, $s$- and $Wt$-channels). We find  improved limits with respect to previous results (in most cases of almost one order of magnitude) and extend the analysis to include four-quark operators.  We  find that  new electroweak physics is constrained to live above  an energy   scale between 430 GeV and 3.2  TeV (depending on the form of its contribution); strongly interacting new physics is bounded by scales higher than 1.3 or 1.5 TeV (again depending on its contribution).
\end{abstract}
\maketitle
%
\vskip1.5em

\section{Motivations and notation} 
Precision studies of the interaction vertices of the top quark to the gauge bosons provide an important tool in testing the standard  model (SM) and searching for new physics contributions. Currently available measurements from the Tevatron and the LHC allow to set limits on possible deviations in the values of the top quark couplings from their SM values. A model independent framework to study these anomalous couplings is effective field theory, where the modifications are encoded into the coefficients of a set of higher dimensional operators that parametrize the effects of new physics at low energy.  The language of effective field theory is very helpful in identifying the source of top quark anomalous couplings and guiding us toward the identification of possible new physics contributions.

The top quark  has both strong and electroweak (EW) interactions. While all interactions come together in collider physics, it is possible to separate in most processes the EW from the strong part so that the anomalous vertices can be discussed separately. 
In this paper we concentrate on the study of the $Wtb$ vertex; the strong interaction $Gt\bar t$ vertex was recently studied in \cite{us}---the results of which are  here   reported. Possible deviations in the interaction between the top quark and the neutral bosons $Z$  and 
$\gamma$---they enter the associated productions $t\bar tZ$ and $t\bar t\gamma$---are left out because still poorly measured.

\subsection{Effective $Wtb$ vertex} 

Following the literature~\cite{AguilarSaavedra:2008zc}, we write  the effective lagrangian  that describes, after EW symmetry breaking, the most general $Wtb$ vertex as
\be \label{wtbv}
{\cal L}_{Wtb}=-\frac{g}{\sqrt{2}}\bar b\gamma^\mu (V_L P_L+ V_R P_R) t W_\mu^-
-\frac{g}{\sqrt{2}}\bar b \frac{i\sigma^{\mu\nu}q_\nu}{m_W}
(g_L P_L+ g_R P_R)t W_\mu^-+\hc \, ,
\ee
where $g$ is the $SU(2)_L$ gauge coupling, $P_{L,R}$ the chiral projectors $(1\pm \gamma_5)/2$ and $\sigma^{\mu\nu}=i[\gamma^\mu,\gamma^\nu]/2$. The 
coefficients  $V_L$, $V_R$, $g_L$ and $g_R$ are, in general, complex dimensionless constants. In this work we will restrict ourselves to the CP-conserving case and these couplings are taken to be real. 
In the SM the $Wtb$ vertex in \eq{wtbv} reduces at the tree level to the Dirac vertex with $V_L =  1$ (after mass diagonalization, $V_L = V_{tb}\simeq 1$) and we take it to be positive. Corrections to $V_L$, as well as the others non-zero anomalous couplings $V_R$, $g_L$, $g_R$ can be generated by new physics. 

 If the physics beyond the SM lies at an energy scale $\Lambda$ that is larger than the  EW scale $\upsilon$, then we can parametrize its effects via higher dimensional operators respecting the SM symmetries in a series  suppressed by inverse powers of the scale $\Lambda$. The leading contributions arise at dimension six and can be written as an expansion of local operators $\{\hat O_k\}$
\be 
{\cal L}_{\mbox{\scriptsize SM}}^{\mbox{\scriptsize dim}\, 6}=\sum_k \frac{c_k}{\Lambda^2}\hat O_k \, ,
\ee
where $c_k$ are dimensionless coefficients.

The complete list of independent dimension six SM effective operators is reported in \cite{Grzadkowski:2010es} of which we follow the notation. The subset of operators that contributes to the anomalous couplings in \eq{wtbv} consists of the following operators
\bea \label{effops}
\hat O^{(3)}_{\varphi q}&=&(\varphi^\dagger i\overleftrightarrow{D}_\mu^I\varphi) (\bar q_L\sigma^I\gamma^\mu q_L)\nn\\
\hat O_{\varphi tb}&=&i(\tilde\varphi^\dagger D_\mu\varphi) (\bar t_R\gamma^\mu b_R)\nn\\
\hat O_{tW}&=&\bar q_L\sigma^{\mu\nu}\sigma^It_R\tilde \varphi W_{\mu\nu}^I\nn\\
\hat O_{bW}&=&\bar q_L\sigma^{\mu\nu}\sigma^Ib_R \varphi W_{\mu\nu}^I
\eea
where $q^T= (t\,\,b)$, $\tilde \varphi=i\sigma^2\varphi^*$ and $D_\mu \varphi=\partial_\mu \varphi +igW_\mu^I\frac{\sigma^I}{2}\varphi+ig'Y_\varphi B_\mu\varphi$.
After EW symmetry breaking $\varphi=(0 \,\,\upsilon +H/\sqrt{2})$, $\upsilon=246$ GeV, we can express the anomalous couplings in terms of the effective field theory coefficients as follows
\bea 
V_L&=&V_{tb}+c^{(3)}_{\varphi q}\frac{\upsilon^2}{\Lambda^2} \simeq 1+c^{(3)}_{\varphi q}\frac{\upsilon^2}{\Lambda^2}\nn\\
V_R&=&\frac{1}{2}c_{\varphi tb}\frac{\upsilon^2}{\Lambda^2}\nn\\
g_R&=&\sqrt{2}\,c_{tW}\frac{\upsilon^2}{\Lambda^2}
\nn\\
g_L&=&\sqrt{2}\,c_{bW}\frac{\upsilon^2}{\Lambda^2} \, .
\eea 
There is another operator that enters in the processes we consider, it is the four fermion operator
\be 
\hat O_{qq'}^{(3)}=\bar q_L\gamma^{\mu}\sigma^I q_L  \bar q'_L\gamma^{\mu}\sigma^I q'_L\label{4f} \, ,
\ee
where $q'^T= (u\,\,d)$ and, as before, $q^T= (t\,\,b)$. It does not give a direct contribution to the anomalous couplings but its interplay with the other operators modifies the limits. For this reason it must be included and its effect parametrized by a new coefficient
\be
C_{4f} =  c_{qq'}^{(3)} \frac{v^2}{\Lambda^2} \, , \label{def}
\ee
which can be further identified by taking $c_{qq'}^{(3)} = 2\pi$ with the usual convention of writing four-quark operators as
\be
\frac{2\pi}{\Lambda^2} \, \bar \psi_L \gamma^\mu \psi_L \, \bar \psi_L \gamma_\mu \psi_L \, ,
\ee
where the coefficient $2\pi$ represents the strength of a strongly interacting new-physics sector---the integration of which gives rise to the effective operator.

The effect of these operators on the top-quark EW anomalous couplings can be best studied in two processes:  top decay (by means of  the $W$ polarizations) and  single-top production.

\subsection{Effective $Gt\bar t$ vertex} 

The effective vertex for the $Gt\bar t$ interactions can be written as
\be
{\cal L}_{Gt\bar t}= - g_s \bar t \left[ \gamma^\mu F_1(q^2)  +  \frac{i \sigma^{\mu\nu}q_\nu}{2 m_t} F_2(q^2) \right]   \, G_\mu \,t \, , \label{int}
\ee
where $g_s$ is the strong $SU(3)_C$ coupling constant, $G_{\mu}= T_A G_{\mu}^A$ is the gluon field, $T_A$ are the $SU(3)_C$ group generators, $q^\mu$ is the momentum carried by the gluon, $t$ denotes the top quark field. The interaction in \eq{int} is the most general after assuming that the vector-like nature of the gluon-top quark vertex  is preserved by the underlying dynamics  giving rise to the composite state. 

The leading contributions come from the following two higher dimensional operators:
\be\label{O}
\hat O_1 = g_s \frac{C_1}{m_t^2} \bar t \gamma^\mu T_A t D^\nu G^A_{\mu\nu} \quad \mbox{and} \quad \hat O_2= g_s \frac{C_2\upsilon}{2 \sqrt{2} m_t^2 } \bar t \sigma^{\mu\nu} T_A t G^A _{\mu\nu}
\ee
where $D^\nu G^A_{\mu\nu}=\partial^\nu  G^A_{\mu\nu} + g_s f^A{}_{BC}  G^{\nu B}  G^C_{\mu\nu}$, $G^A_{\mu\nu}=\partial_\mu G_\nu^A-\partial_\nu G_\mu^A +g_s f^A{}_{BC}  G_\mu^B G_\nu^C$ is the gluon field strength tensor, $f^A{}_{BC}$ are the $SU(3)_C$ structure constants. In \eq{O} the operator $\hat O_1$ and $\hat O_2$ are, respectively, of dimension six and five. We limit ourselves to the $CP$-conserving case and the dimensionless coefficients $C_1$ and $C_2$ are taken to be real. The operator $\hat O_1$ gives the leading $q^2$ dependence to $F_1$ while $\hat O_2$ gives the $q^2$-independent term of $F_2$:
\be 
\label{f12}
F_1(q^2)=1+C_1 \frac{q^2}{m_t^2} +\ldots\qquad\mbox{and}\qquad F_2(0)=\sqrt{2}\, C_2 \frac{\upsilon}{m_t} \,.
\ee

Operators of higher dimensions can in general contribute---they gives further terms in the expansion of the form factors---but their effect is very much suppressed.
The form of the coefficients in front of the operators in \eq{O} is conventional and dictated, in our case, by the analogy with the electromagnetic form factors. In addition, the operator $\hat O_2$ is written for convenience with an extra factor $\upsilon/m_t$ because it can be thought as coming, after EW symmetry breaking, from a dimension six $SU(3)_C\times SU(2)_L\times U(1)_Y$ gauge invariant operator that includes the Higgs boson field.

These vertices were analyzed (most recently in \cite{us} where references to all related works can be found)  by means of the $t\bar t$ production cross section and spin correlations. The following limits  are found:
\be
-0.008 \leq C_1 \leq  0.015 \qquad \mbox{and} \qquad   -0.023 \leq C_2 \leq 0.020 \, .
\ee
These limits give a direct bound on the  magnetic moment $\mu$ and the RMS radius $\langle \vec r^2 \rangle$, traditionally used to characterize the size of extended objects. They correspond to 
\be
\sqrt{\langle \vec r^2 \rangle} <  4.6 \times 10^{-4} \; \mbox{fm} \quad \mbox{(95\% CL)} \quad 
 \mbox{and} 
-0.046 < \mu -1 < 0.040 \quad \mbox{(95\% CL)} \, . \label{mu}
\ee

We  take the coefficients $C_1$ and $C_2$ to be zero when studying EW anomalous couplings---they only enter one channel of the single-top production---and assume that the strong interaction of the top quark follows the SM in all production processes.

\section{Methods}
\label{sec:mc}

In order to study the effect of anomalous couplings on top decay rates and single top production cross sections at the LHC and Tevatron, we  first use {\sc FeynRules}~\cite{Christensen:2008py} to implement our model, which has been defined to be the SM with the addition of the  effective operators in \eq{effops} and \eq{4f}. The dependence on these operators is encoded in the coefficients $c_i = V_L, V_R, g_L, g_R, C_{4f}$. 

{\sc FeynRules} provides the Universal FeynRules Output (UFO) with the Feynman rules of the model. The UFO is then used by {\sc MadGraph 5} \cite{madgraph} (MG5)  to compute the   branching ratios  in the decay rates and production cross sections, which we denote by $F^{\mbox{\scriptsize MG5}}_{L,0}$ and $\sigma_{\mbox{\scriptsize MG5}}$, respectively. 

 MG5 computes the square of the amplitude for each single channel.
The partonic level result thus obtained can be compared with the partonic experimental cross section that is extracted by the experimental collaborations.

We compute, using MG5, the top-quark decay width and the single-top production cross section  varying the  values of  the anomalous couplings $c_i$ in a range that goes from $-2$ to $2$, except for $V_L$ that is  varied only for positive values. These different values of branching ratios $F^{\mbox{\scriptsize MG5}}_{L,0}(c_i)$ and cross sections $\sigma_{\mbox{\scriptsize MG5}}(c_i)$ are used to obtain limits on the coefficients  by comparing the MG5 computation with the  measured cross sections and helicity fractions at the LHC at the  center-of-mass (CM) energies $\sqrt{s}= 7$ and 8 TeV, as discussed in the next section.
By proceeding in the same way, we have also computed the rate and production cross section at the Tevatron and compared it with the measured cross section at the CM energy $\sqrt{s} = 1.98$ TeV. 

Analytical expressions for the amplitudes that we study numerically can be found in~\cite{Zhang:2010dr}.

\subsection{Statistical analysis}

To obtain $95\%$ confidence level (CL) limits on the coefficients $c_i$ from the production cross sections, 
we use the quantity $\Delta\sigma_{\mbox{\scriptsize exp}}$, 
defined to be the difference between the central value of the measured cross section $\bar \sigma_{\mbox{\scriptsize exp}}$ 
and that of the SM theoretical value $\bar \sigma_{\mbox{\scriptsize th}}$:
\be
\Delta \sigma_{\mbox{\scriptsize exp}} = \bar \sigma_{\mbox{\scriptsize exp}} - \bar \sigma_{\mbox{\scriptsize th}} \, .
\ee

The uncertainty on $\Delta\sigma_{\mbox{\scriptsize exp}}$ is given by summing in quadrature the respective uncertainties:
\be
\sqrt{(\delta\sigma_{\mbox{\scriptsize exp}})^2 + (\delta\sigma_{\mbox{\scriptsize th}})^2}\,.
\ee

Using the cross sections $\sigma_{\mbox{\scriptsize MG5}}(c_i)$ calculated with MG5 
we compute the value of the cross section coming from new physics as:
\be
\Delta \sigma_{\mbox{\scriptsize MG5}}(c_i) =  \sigma_{\mbox{\scriptsize MG5}}(c_i) - \sigma_{\mbox{\scriptsize MG5}}(0) \, .
\ee

The quantity $\Delta \sigma_{\mbox{\scriptsize MG5}}(c_i)$ represents the contribution to the cross section coming from the
interference between SM leading order and new physics diagrams as well as terms quadratic in the anomalous couplings.

Values of $c_i$  for which
$\Delta \sigma_{\mbox{\scriptsize MG5}}(c_i)$ is more
than two standard deviations from $\Delta \sigma_{\mbox{\scriptsize exp}}$, namely:
\be
\Delta \sigma_{\mbox{\scriptsize MG5}}(c_i) > \Delta \sigma_{\mbox{\scriptsize exp}} + 2 \, \sqrt{(\delta\sigma_{\mbox{\scriptsize exp}})^2 + (\delta\sigma_{\mbox{\scriptsize th}})^2} \, , \label{delta1}
\ee
or
\be
\Delta \sigma_{\mbox{\scriptsize MG5}}(c_i) < \Delta \sigma_{\mbox{\scriptsize exp}} - 2 \, \sqrt{(\delta\sigma_{\mbox{\scriptsize exp}})^2 + (\delta\sigma_{\mbox{\scriptsize th}})^2} \, , \label{delta2}
\ee
are considered excluded at 95\% CL.

To obtain limits from the branching ratios in top decay, as measured by the $W$-boson helicity fractions, a similar procedure is used.
The experiments measure the three fractions $F_0$, $F_L$ and $F_R$ (see below for the definition). 
Being extracted from the same data set and from the same observable, these fractions are not independent. 
$F_R$ is constrained to be equal to $1-F_0-F_L$ and is therefore not considered in the limit extraction. 
In addition, there is a correlation factor $\rho$ different from 0 between the measured values of $F_0$ and $F_L$. 

The helicity fractions $F_0^{\mbox{\scriptsize MG5}}(c_i)$ and $F_L^{\mbox{\scriptsize MG5}}(c_i)$ are computed within MG5, after adjusting for higher order QCD corrections,
and  compared to the measured values $F_0^{\mbox{\scriptsize exp}}$ and $F_L^{\mbox{\scriptsize exp}}$. 
The only uncertainties we consider are the experimental uncertainties 
$\delta F_0^{\mbox{\scriptsize exp}}$ and $\delta F_L^{\mbox{\scriptsize exp}}$, 
being the theoretical ones negligible. 

To extract the limits, a likelihood function is defined as:
\be
L(c_i) \propto e^{\displaystyle
  -\frac{1}{2(1-\rho^2)}
  \left[
    \chi_0^2(c_i) + \chi_L^2(c_i) - 2\rho \cdot \chi_0(c_i) \cdot \chi_L(c_i)
  \right]
},
\ee
where:
\be
\chi_0(c_i) = \frac{F_0^{\mbox{\scriptsize exp}} - F_0^{\mbox{\scriptsize MG5}}(c_i)}{\delta F_0}
\quad \mbox{and} \quad
\chi_L(c_i) = \frac{F_L^{\mbox{\scriptsize exp}} - F_L^{\mbox{\scriptsize MG5}}(c_i)}{\delta F_L},
\ee
and the constrain $\int L(c_i) dc_i = 1$ is imposed.

Then, the quantity $L^{95\%}$ is defined as:
\be
\int_{L(c_i)>L^{95\%}} L(c_i) dc_i = 0.95
\ee
and values of $c_i$ for which $L(c_i)<L^{95\%}$ are considered excluded at 95\% CL.

\section{Results} 
\label{sec:r}

\begin{figure}[t!]
\begin{center}
\includegraphics[width=4in]{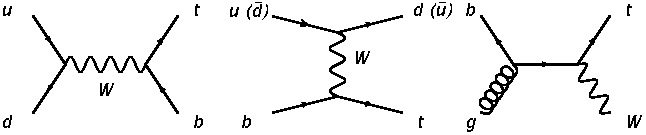}
\caption{\small  Tree-level diagrams for, from left to right, $s$-, $t$- and $Wt$-channel.
\label{channels}}
\end{center}
\end{figure}
\begin{figure}[t!]
\begin{center}
\includegraphics[width=4in]{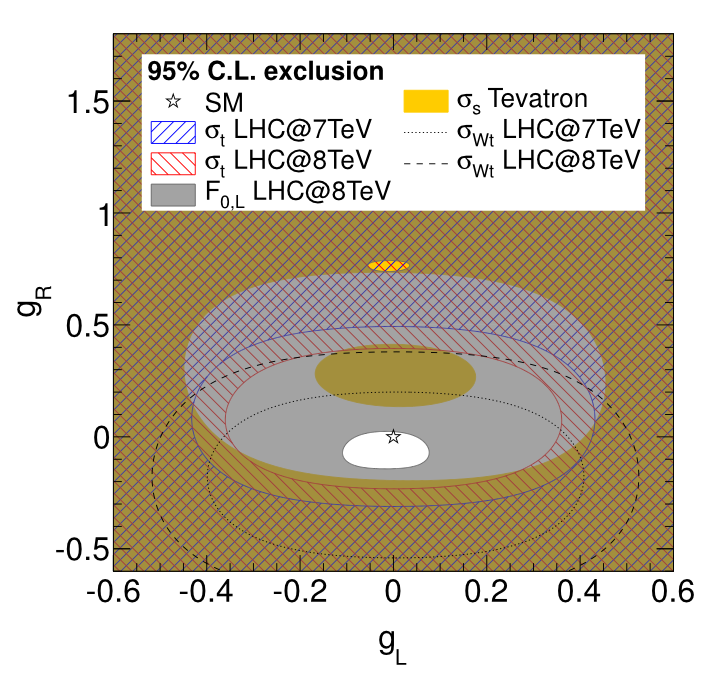}
\caption{\small  95\% C.L. exclusion limits on the coefficients $g_R$ and $g_L$ ($V_L=1$, $V_R=C_{4f}=0$).
The full yellow, striped red, striped blue and shaded gray areas indicate the excluded regions from the $s$- and $t$-channel production cross-sections and from the helicity fractions. The area outside the dashed (dotted) ellipses is excluded by the $Wt$-channel cross-section measurement at the LHC at 8 TeV (7 TeV).
Region of allowed values: $-0.109 \leq g_L \leq 0.076$ and $-0.142 \leq g_R \leq 0.024$\label{gRgL}}
\end{center}
\end{figure}

Top-quark decay  and single-top production  are the two processes where we can  directly probe the top coupling to the weak bosons.   

The $Wbt$ vertex in top-quark decay is best studied by means of the helicity fractions $F_0$ and $F_L$ defined, respectively, as the ratios between the rates of polarized decay of the top quark into zero and left-handed $W$ bosons and the total decay width.

In the SM they are found, neglecting $m_b$, to be
\be
F_0 = \frac{m_t^2}{m_t^2 + 2 m_W^2} \simeq 0.7 \quad \mbox{and} \quad
F_L = \frac{2 m_W^2}{m_t^2 + 2 m_W^2} \simeq 0.3 \, ;
\ee
the helicity fraction into right-handed $W$ bosons is vanishingly small. These branching ratios  receive corrections from the operators $\hat O_{tW}$ and $\hat O_{bW}$. The SM values at the next-to-next-to-leading order (NNLO) order in QCD are computed in~\cite{Gao:2012ja,Czarnecki:2010gb}.

The best current experimental results are given by CMS~\cite{heli} for data at $\sqrt{s} = 8$ TeV (integrated luminosity $19.6$ fb$^{-1}$) as
\bea
F_0 &=&0.659 \pm 0.015  \;\mbox{(stat.)} \pm 0.023  \;\mbox{(syst.)}  \nn \\
F_L&=&0.350 \pm 0.010  \;\mbox{(stat.)}  \pm 0.024 \;\mbox{(syst.)} \, ,
\eea
with a correlation coefficient $\rho = 0.95$.

The single-top production occurs through $t$-, $s$- and ${Wt}$-channel (see Fig.~\ref{channels}). The SM NNLO cross sections have been computed in \cite{Kidonakis:2011wy}, \cite{Kidonakis:2010tc} and \cite{Kidonakis:2010dk}, respectively. The corresponding experimental cross sections $\sigma_t$, $\sigma_s$ and $\sigma_{Wt}$ are available  from the LHC and Tevatron data and we utilize the following results (in which the error includes both statistical and systematic contributions): 
\bea
\sigma_t &=& 67.2 \pm 6.1 \; \mbox{pb} \quad \mbox{(LHC@7 TeV) \cite{Chatrchyan:2012ep}} \nn \\
\sigma_t &=& 83.6 \pm 7.7 \; \mbox{pb}  \quad \mbox{(LHC@8 TeV) \cite{Khachatryan:2014iya}} \, ,
\eea
for single-top production in the $t$-channel (integrated luminosities of 1.17 and 1.56 fb$^{-1}$ for, respectively,  muon and electron final states at 7 TeV and 19.7 fb$^{-1}$ at 8 TeV),
\be
\sigma_s = 1.29^{+0.26}_{-0.24}  \quad \mbox{(CDF+D0@1.98) \cite{CDF:2014uma}} \, ,
\ee
for the $s$-channel (integrated luminosity $7.5$ pb$^{-1}$)  and 
\bea
 \sigma_{Wt} &= &16^{+5}_{-4}  \; \mbox{pb}  \quad \mbox{(LHC@7 TeV) \cite{Chatrchyan:2012zca}} \nn \\
 \sigma_{Wt} &= & 23.4^{+5.5}_{-5.4} \; \mbox{pb}  \quad \mbox{(LHC@8 TeV) \cite{Chatrchyan:2014tua}} \, ,
\eea
for the  $Wt$-channel (integrated luminosities 4.9 fb$^{-1}$ at 7 TeV and 12.2 fb$^{-1}$ at 8 TeV) .

\begin{figure}[t!]
\begin{center}
\includegraphics[width=4in]{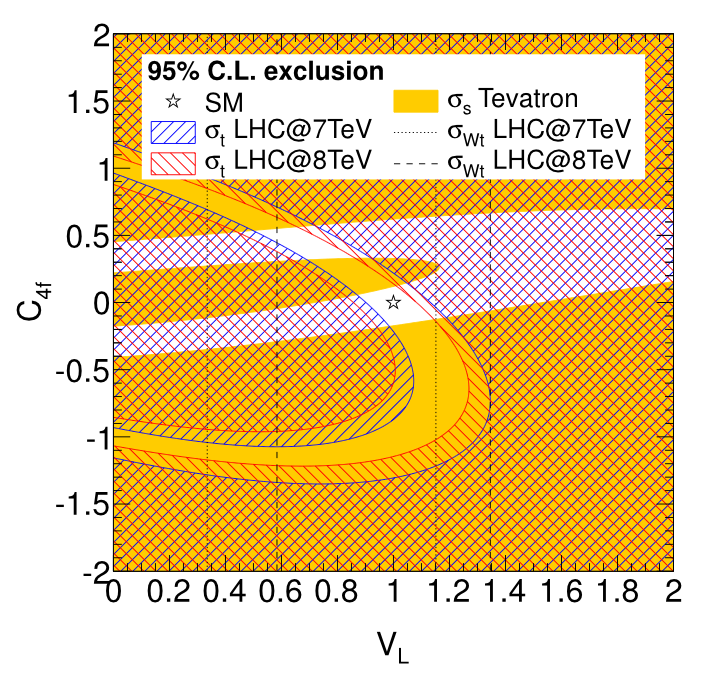}
\caption{\small  95\% C.L. exclusion limits on the coefficients $C_{4f}$ and $V_L$ ($g_L=g_R=V_R=0$).
The full yellow, striped red, striped blue and shaded gray areas indicate the excluded regions from the $s$- and $t$-channel production cross-sections and from the helicity fractions. The areas outside the dashed (dotted) vertical lines are excluded by the $Wt$-channel cross-section measurement at the LHC at 8 TeV (7 TeV). Region of allowed values: $0.584 \leq V_L \leq 1.145$ and $-0.037 \leq C_{4f} \leq 0.120$.\label{C4fVL}}
\end{center}
\end{figure}

\begin{figure}[h!]
\begin{center}
\includegraphics[width=4in]{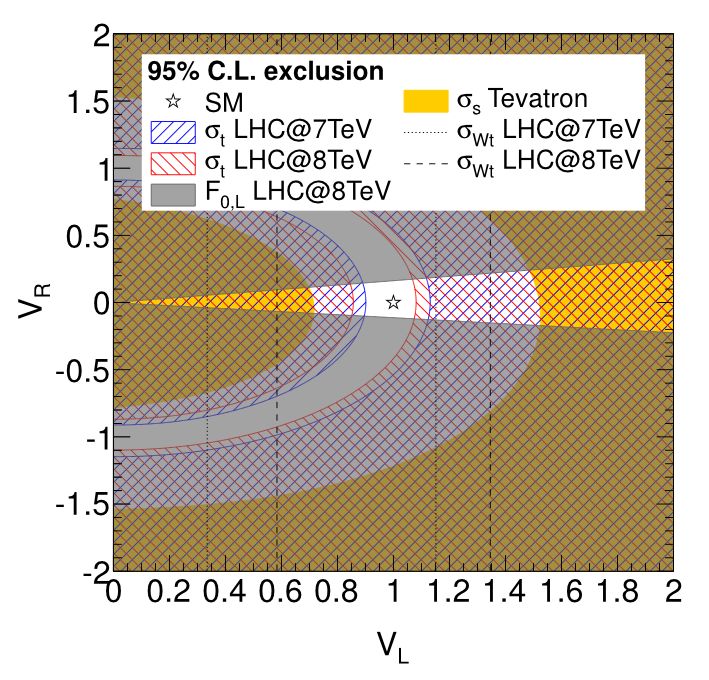}
\caption{\small  95\% C.L. exclusion limits on the coefficients $V_R$ and $V_L$ ($g_L=g_R=C_{4f}=0$).
The full yellow, striped red, striped blue and shaded gray areas indicate the excluded regions from the $s$- and $t$-channel production cross-sections and from the helicity fractions. The areas outside the dashed (dotted) vertical lines are excluded by the $Wt$-channel cross-section measurement at the LHC at 8 TeV (7 TeV).  Region of allowed values: $0.891 \leq V_L \leq 1.081$ and $-0.121 \leq V_R \leq 0.173$.
\label{VRVL}}
\end{center}
\end{figure}

We study the effect of the operators in \eq{effops} and \eq{4f} by varying the coefficients of two of them at the time and we derive the related limits following the statistical analysis described in the previous section.  The results are depicted in Figs.~\ref{gRgL}, \ref{C4fVL} and \ref{VRVL}.

The determination of the limits on the coefficients $g_R$ and $g_L$ is dominated by the top-quark decay but the single-top production cross section is   useful in eliminating larger values (the small area on top of Fig.~\ref{gRgL}). Accordingly, only the lower region is allowed and 
\be
-0.109 \leq g_L \leq 0.076\quad \mbox{and} -0.142 \leq g_R \leq 0.024 \, .
\ee

The interplay  among the various channels of single-top production is crucial in delimiting the allowed region for the parameters $C_{4f}$ and $V_L$ in Fig~\ref{C4fVL}. Unfortunately it is not sufficient in completely delimiting the allowed range of the coefficients to a single region.  The $Wt$-channel is not sensitive to $V_R$ and $C_{4f}$. Accordingly, the final bound is weaken when both coefficients are allowed to vary simultaneously and we find
\be
0.584 \leq V_L \leq 1.145\quad \mbox{and}  -0.037 \leq C_{4f} \leq 0.120 \, .
\ee
A future improvement in the $Wt$-channel measurement at the LHC could be instrumental in delimiting the range to a single region.

Finally, it is the interplay between top decay and the $t$-channel of single-top production that provides the limit on the coefficients $V_R$ and $V_L$ (see Fig.~\ref{VRVL}).  We find
\be
0.891 \leq V_L \leq 1.081 \quad \mbox{and}   -0.121 \leq V_R \leq 0.173 \, .
\ee 

Notice that the coefficient $V_R$ is very much constrained by flavor physics~\cite{Vignaroli:2012si} from which it should be 
\be
-0.0004 \leq V_R \leq 0.0013 \, .
\ee 
 When this limit is included, Fig.~\ref{VRVL} should be read only along the line $V_R\simeq 0$ with a slightly improved bound 
 \be 
 0.902  \leq V_L \leq 1.081 \, . 
 \ee

Our results are summarized in Table~\ref{limits} in the case in which the various anomalous couplings are turned on one at the time. 

\begin{table}[ht!]
\begin{center}
\caption{Limits (95\% CL) on the coefficients $c_i = V_L, V_R, g_L, g_R, C_{4f}$ and $C_{1,2}$ when they are varied independently of each other and  energy scale of the corresponding effective operators.}
\label{limits}
\vspace{0.2cm}
\begin{tabular}{|c|c|}
\hline
\qquad $-0.142\leq g_R \leq 0.023$ \qquad  &\qquad $|g_R| \leq 0.142$,  \quad $\Lambda\gtap 780$ GeV \qquad \cr
\hline
\qquad $-0.081 \leq g_L \leq 0.049$\qquad  &\qquad $|g_L| \leq 0.081$, \quad $\Lambda\gtap $ 1 TeV \qquad \cr
\hline
\qquad $0.902 \leq V_L \leq 1.081$ \qquad & \qquad $|V_L-1| \leq 0.098$, \quad   $\Lambda\gtap $ 790 GeV \qquad \cr
\hline
\qquad $-0.112 \leq V_R \leq 0.162$\qquad  & \qquad $|V_R| \leq 0.162$, \quad   $\Lambda\gtap $ 430 GeV \qquad \cr
\hline
\qquad $ -0.036 \leq C_{4f} \leq  0.025$ \qquad & \qquad $|C_{4f}| \leq 0.036$, \quad  $\Lambda\gtap $ 3.2 TeV\qquad  \cr
\hline
\hline
\qquad $-0.008 \leq C_1 \leq 0.015$\qquad  & \qquad $|C_1| \leq 0.015$, \quad   $\Lambda\gtap $ 1.3 TeV \qquad \cr
\hline
\qquad $ -0.023 \leq C_2 \leq  0.020$ \qquad & \qquad $|C_2| \leq 0.023$, \quad  $\Lambda\gtap $ 1.5 TeV\qquad  \cr
\hline
\end{tabular}
\end{center}
\end{table}

The limits found can be  interpreted  in terms of the energy scale of the effective operators in \eq{effops} (see second column of Table~\ref{limits}).
 By inspection, we see that all EW limits  are around $10^{-1}$ which translates into a characteristic scale  $\Lambda \simeq 700$ GeV (actually from 430 GeV to 1 TeV, depending on the contribution) except for the four-quark interaction which has $\Lambda \simeq 3.2$ TeV, if we follow the definition in \eq{def}. Above these limits,  there is still room for new physics. These results  can be compared with the strong sector interactions of the top quark where the corresponding energy scale is $\Lambda \simeq 1.3$ and 1.5 TeV~for, respectively, the operators $\hat O_1$ and $\hat O_2$ in \eq{O}~\cite{us}.

\section{Discussion} 

The previous status of  limits on anomalous top quark gauge couplings was the following. 

Constraints on the $Wtb$ vertex from Tevatron and early LHC data  were reported in \cite{AguilarSaavedra:2011ct}. The bounds  were obtained, as in this paper,  by combining the experimental measurements of $W$-boson helicity fractions and single-top production and by varying two operators at the time. 

More recently, the measurement of the $W$-boson helicity fractions has been substantially improved by ATLAS and CMS in \cite{ATLAS:2013tla}, \cite{Chatrchyan:2013jna}  and \cite{CMS:exa} and limits on the coefficients $g_R$ and $g_L$ were extracted by the experimental collaborations based on these new measurements.
Concerning the single top production, many detailed results have been released by both Tevatron and LHC experiments~\cite{Chatrchyan:2012ep,Khachatryan:2014iya,CDF:2014uma,Chatrchyan:2012zca,Chatrchyan:2014tua}, as reported in the previous section. 

We confirm and improve the above limits by means of the most recent data set at 8 TeV by CMS~\cite{heli} and the combined use of the helicity fractions and single-top production cross sections. The improvement is substantial, in most cases of almost one order of magnitude. 
 As expected, the simultaneous presence of more than one anomalous coupling  weaken the limits.  The result for the four-fermion operator (\ref{4f}), and its impact on the determination of the coefficient $V_L$  is new. 
 
 \begin{acknowledgments}
MF is associated to SISSA.   The work of AT 
was supported by the S\~ao Paulo Research Foundation (FAPESP) under grants 2011/11973-4 and 2013/02404-1.
\end{acknowledgments}


\end{document}